\documentstyle[preprint,aps]{revtex}
\begin{document}

\draft
\title
{Anisotropic strains and magnetoresistance of 
La$_{0.7}$Ca$_{0.3}$MnO$_{3}$}
\author
{T. Y. Koo, S. H. Park, K.-B. Lee, and Y. H. Jeong\cite{jeong}}
\address{Department of Physics, 
Pohang University of Science and Technology, Pohang, Kyungbuk 790-784, 
S. Korea}
\maketitle

\begin{abstract} 
Thin films of perovskite manganite La$_{0.7}$Ca$_{0.3}$MnO$_{3}$ were 
grown epitaxially on 
SrTiO$_3$(100), MgO(100) and LaAlO$_3$(100) substrates by the pulsed laser 
deposition method. 
Microscopic structures of these thin film samples as well as a bulk sample 
were fully determined by x-ray 
diffraction measurements. The unit cells of the 
three films have different shapes, i.e., contracted tetragonal, 
cubic, and elongated tetragonal for SrTiO$_3$, MgO, and LaAlO$_3$ cases, 
respectively, while the unit cell of the bulk is cubic.   
It is found that the samples with cubic unit cell show smaller
peak magnetoresistance than the noncubic ones do.  The present 
result demonstrates that the magnetoresistance of La$_{0.7}$Ca$_{0.3}$MnO$_{3}$  
can be controlled by lattice distortion via externally imposed strains. 
\end{abstract}
\vskip2pc
\narrowtext
 
The compounds of mixed valence manganite with the perovskite 
structure, La$_{1-x}$D$_x$MnO$_3$ with D for a divalent metal ion, have 
attracted renewed attention due to their anomalously large negative 
magnetoresistance (MR) effects \cite{jin1,kusters,chahara,helmolt}.  
While these manganites are paramagnetic insulators at high 
temperatures, a compound with $0.2<x<0.4$ shows a simultaneous 
appearance of metallic conduction and ferromagnetism below a certain 
ordering temperature $T_c$, around which the large MR effect is found
\cite{jonker,wollan}. (Resistance at zero field reaches a maximum at 
$T_{p}\,\approx\,T_{c}$.) As a function of composition high $T_{c}$ 
and large values of MR are usually achieved in a 
sample with $x\approx$0.3, which is thus of particular interest. 
Zener's double exchange model (DEM)
accounts for ferromagnetic ordering in this system in terms of the itinerancy of 
charge carriers, which is characterized by the Mn to Mn hopping matrix 
element, $t=t_{0} \cos
\frac{\theta}{2}$ where $\theta$ is the angle between the core spins 
at nearest Mn sites and $t_{0}$ is the bare matrix element \cite{zener}. 
The 
$\theta$ dependence of $t$ gives rise to negative magnetoresistance. 
However, recent theories have emphasized the importance of the polaron effect 
associated with dynamic Jahn-Teller distortion of Mn$^{3+}$ \cite{millis}. 
Band structure calculations for the D=Ca system indicate the 
half-metallic nature and strong Mn-O hybridization for the majority 
band; this unusual band structure was suggested as the reason for 
magnetoresistive behaviors \cite{band}.

Due to numerous potential applications of large MR materials, 
such as magnetic recording, sensors, 
and switching devices, the major research focus of mixed valence 
manganites has been on increasing the value of MR 
and shifting the MR peak above room temperature. 
However, tailoring 
the materials for optimum conditions in the sensitivity and temperature 
variation of the magnetoresistive response would require 
the basic understanding of transport phenomena in the system.
In this regard, it would be of value to carefully examine their properties 
as a function of lattice strain.
The DEM, for example, naturally predicts a dependence of magnetic 
and transport properties on $t$, which in turn would depend on lattice 
constant. Furthermore, Pickett and Singh's calculation \cite{band} 
identified lattice distortion as an important factor in determining 
the band structures of the system.

Previous studies of the lattice effects on magnetotransport properties 
of perovskite manganites include two cases: first,
the average radius of trivalent A-site ions was changed by mixing two 
kinds of ions of different radius, for 
example, La and Pr, while keeping the divalent ion concentration $x$, 
which determines the carrier density, fixed. Then the variation of 
the trivalent A-site radius would cause 
internal stresses (chemical pressure) in the system \cite{jin2,hwang1}.
When this average radius is reduced, it turned out that $T_p$ is 
decreased and the magnitude 
of MR is enhanced.
While Hwang et al. interpreted these results in terms of Mn-O-Mn bond 
angle variation, it is quite obvious that the whole effect cannot 
be attributed solely 
to the geometrical lattice effects and the chemical origin must be 
taken into account \cite{goodenough}. 
The second kind of experiments 
were conducted by applying an external hydrostatic 
pressure \cite{hwang2,khazeni,neumeier}.
The application of hydrostatic pressure increases 
$T_p$ monotonically and strongly suppresses the 
magnitude of MR. In these cases, the pressure 
range was limited and no strain data were reported.

Thus, the current situation calls for systematic investigations on 
the pure lattice effects on the magnetotransport properties of 
perovskite manganites in the absence of chemical interference; in 
particular, the effects 
of anisotropic strains would be of great interest.
For this purpose, we utilized the fact that 
the lattice parameters of films are different from that of a bulk and
can be varied by depositing films on various substrates. We were 
able to vary the {\it lattice symmetry} as well as the lattice constants of 
La$_{0.7}$Ca$_{0.3}$MnO$_{3}$ (LCMO) thin films. 
For the first time to our knowledge it is
revealed that crystal symmetry is an important factor in controlling 
magnetotransport properties of LCMO.

LCMO bulk samples were prepared by the standard 
solid state reaction technique from 99.9\% purity oxide and carbonate 
powders. After repeated grinding and calcining at 1000$^\circ$C for 24 h,
cold isostatic pressed pellets were sintered at 1350$^\circ$C for 5 h 
in air.
The X-ray diffraction analysis of the pellets showed the single phase 
perovskite structure of cubic symmetry with lattice constant 
a$\,=\,3.867$ \AA. 
Using these pellets as a target, thin films of LCMO were grown on 
SrTiO$_3$(100), MgO(100), and LaAlO$_3$(100) substrates by the pulsed 
laser deposition method under the same growth conditions.
Substrate temperature was controlled at 650 $^\circ$C, and oxygen partial 
pressure was maintained at 150 mtorr during the deposition period.
After deposition, the films were left at 650 $^\circ$C for 10 min 
under an oxygen pressure of 1 torr, and then freely cooled to room temperature. 
The typical thickness of the films was approximately 1000 \AA.
Resistance and magnetoresistance of the LCMO samples, in the film and 
bulk form, were measured by the four-probe 
method as a function of temperature ($T$) and magnetic field ($B$).
$T$ was varied from 50 K to 320 K, and $B$ was changed 
up to 1 T.
The direction of the applied field was parallel to that of the probing current. 
Measurements of magnetic 
susceptibility and magnetization were 
also performed using an ac susceptometer and a vibrating 
sample magnetometer, respectively. These measurements verified
the ferromagnetic phase transition in the samples which accompanies the 
resistance maximum \cite{ku}.

4-circle x-ray diffractometry (CuK$\alpha$ source) was used to characterize 
the microscopic structure of the films at room temperature, and 
confirmed their epitaxial nature. 
The diffraction peaks were precisely determined by fitting 
the split K$\alpha$1 and K$\alpha$2 lines with the two-Gaussian functional form.
The lattice constant of elementary perovskite cell along the film/substrate 
normal direction, c$_p$, was obtained from the position of (001) peaks 
of $\theta$-2$\theta$ scans, 
while the in-plane lattice constants, a$_p$$\simeq$b$_p$, were determined 
by analyzing the (203) and (104) off-normal peaks.
Table \ref{xray} summarizes x-ray diffraction results for the films. 
The result for the bulk target, which is cubic, is also included. 
It should be noticed from the table that the lattice symmetry of 
the films changes 
from contracted tetragonal (the SrTiO$_3$ case) to nearly cubic 
(the MgO case) and to elongated tetragonal 
one (the LaAlO$_3$ case) as unit cell volume V$_p$ decreases; 
this behavior is due to the biaxial nature of strains inherent in 
films \cite{kwak,foster}.

With these structural information in hand, we present the results of transport 
measurements. In Fig. \ref{resistance} plotted is 
the resistance at $B$= 0 and 1 T of the three LCMO films, 
normalized to their respective room temperature values, versus $T$. 
As is easily seen from the figure, 
$T_p$, the maximum resistance temperature 
at zero field, shifts to a higher temperature and the peak value 
of resistance decreases with reduction in unit cell volume. 
Apparently this behavior resembles that of a bulk under a hydrostatic 
pressure mentioned above \cite{hwang2,khazeni,neumeier}. However, 
there exists a clear distinction in the MR behaviors; 
in the inset of Fig. \ref{resistance} shown is the MR, defined as 
$[R(0) - R(B=1T)]/R(0)$, of the films.
The striking feature in the figure is that the maximum MR, which
occurs at approximately 15 K below $T_{p}$, 
is not monotonic as a function of unit cell volume. In fact, 
the film with nearly cubic unit cell (one deposited on MgO) 
displays a smaller value in the maximum MR than the other two films. 
This is in sharp contrast to the  bulk behavior 
where the reduction in 
unit cell volume (or increasing hydrostatic pressure) gives rise to
a monotonic decrease of the maximum MR, 
in conjunction with the monotonic $T_p$ rise \cite{hwang2,khazeni}.

To directly expose the correlation between the lattice structure and the 
magnetotransport properties of LCMO, $T_p$ and the maximum MR values 
of both the films and the bulk are plotted against unit cell volume 
in Fig. \ref{cross}. The nonmonotonic dependence of these properties 
on unit cell volume is indeed evident.
It is particularly clearly manifested in the MR 
behavior, where the samples with cubic symmetry (middle two points in 
the figure) show lower values than the noncubic ones (outer two points) do. 
$T_{p}$ also varies nonmonotonically, but in somewhat different fashion. 
$T_p$ increases from the lowest 
value for the sample with the largest cell volume 
(the one on SrTiO$_3$ with tetragonal symmetry), as the volume 
decreases; after reaching a maximum at the bulk sample (with the cubic cell), 
$T_{p}$ starts to decrease. This $T_{p}$ behavior may be interpreted 
as a superposition of two differing tendencies: one tendency is 
related to the volume effect favoring the monotonic 
increase of $T_{p}$ as the volume of unit cell is reduced. Notably 
this volume effect seems to be lacking in MR.
The other tendency is due to the symmetry effect, in accord with the MR 
behavior, favoring the maximum $T_{p}$ for cubic samples.
From these observations one can conclude that the lattice symmetry, in 
addition to the unit cell volume, plays a crucial role in determining 
magnetotransport properties of LCMO. 
The symmetry change seen in the film samples would not arise in the case 
of a cubic bulk 
under hydrostatic pressure where isotropic strains prevail. 

The DEM may offer some qualitative explanation of the present results, 
that is, the bare matrix element $t_{0}$ in 
$t (=t_{0} \cos \frac{\theta}{2}$) is made to include a structural dependence. 
Since it is known that the band width of a perovskite material 
depends on the bond length \cite{harrison} as well as the bond angle \cite{obradors}, 
$t_{0}$ may be expressed as
$t_{0}\,\sim\,\frac{\cos \phi}{d^3}$, where $d$ is the Mn-O bond 
length and $\phi$ is the deviation of the Mn-O-Mn bond angle from 
180$^{\circ}$. While the volume of unit cell is expected to depend more 
directly on $d$, the noncubic nature of symmetry would be closely related to 
the bond angle parameter $\phi$.
However, the symmetry effect in MR may not be accounted for solely in this model. 
It is worthwhile, in this regard, to recall that the band structure 
of the LCMO system and its half metallic nature 
depend sensitively on lattice distortion, while the volume 
variation is of less importance \cite{band}. It is also noteworthy 
that the externally imposed tetragonal distortions lift the orbital degeneracy 
which is essential for the dynamic Jahn-Teller effect  \cite{millis}. 
In summary, our results thus have important implications on the current 
understanding of ferromagnetic manganites, besides demonstrating that
the MR of LCMO may be controlled by external anisotropic strains.

This work was partially supported by KOSEF (96-0702-01-01-3), BSRI of 
POSTECH (96-2438), and 
the Korean Ministry of Science and Technology.

\begin{figure}
\caption{Resistivity of La$_{0.7}$Ca$_{0.3}$MnO$_3$ thin films 
on SrTiO$_3$(100), MgO(100), and LaAlO$_3$(100) is plotted as a function of 
temperature ($T$) and field 
($B$). Resistivity of each sample is normalized to the value at $T$= 317 
K and zero field. 
No appreciable thermal hysteresis was found in the measurements and 
the data obtained in cooling are displayed. In the inset shown is 
magnetoresistance, defined as $[R(0) - R(B=1T)]/R(0)$, of the films.
}
\label{resistance}
\end{figure}

\begin{figure}
\caption{Peak resistance temperature $T_p$ at zero field (circles) and 
maximum MR (squares) are plotted against the unit cell volume V$_p$. 
The solid symbols denote the data for the films, and the empty ones 
do those for the bulk. Note the symmetry change as the volume varies. 
The unit cells of the samples depicted by the middle two points are 
cubic, while the outer two points represent the samples with 
tetragonal symmetry.
The lines are guides for the eye.
}
\label{cross}
\end{figure}


\begin{table}
\caption
{Lattice parameters (a$_p$, b$_p$, c$_{p}$) and unit cell volume 
(V$_{p}$) of LCMO samples obtained from 
x-ray diffraction measurements at room temperature.  The lattice 
parameters represent the elementary perovskite unit cell.  Thin films 
are designated by substrates.}
\label{xray}

\begin{tabular}{|c|c||c|c|c|}
\hline
substrate &  cubic a (substrate, \AA) & a$_p$ $\approx$ b$_p$ 
(LCMO in-plane, \AA) & c$_p$ (LCMO normal, \AA) & V$_p$ (\AA$
^3$) 
\\ \hline
LaAlO$_3$(100) & 3.794 & 3.842 & 3.878 & 57.24 \\ \hline
LCMO Target    &       & 3.867 & 3.867 & 57.83 \\ \hline
MgO(100)       & 4.216 & 3.885 & 3.891 & 58.73 \\ \hline
SrTiO$_3$(100) & 3.895 & 3.921 & 3.845 & 59.11  \\ \hline
\end{tabular}

\end{table}

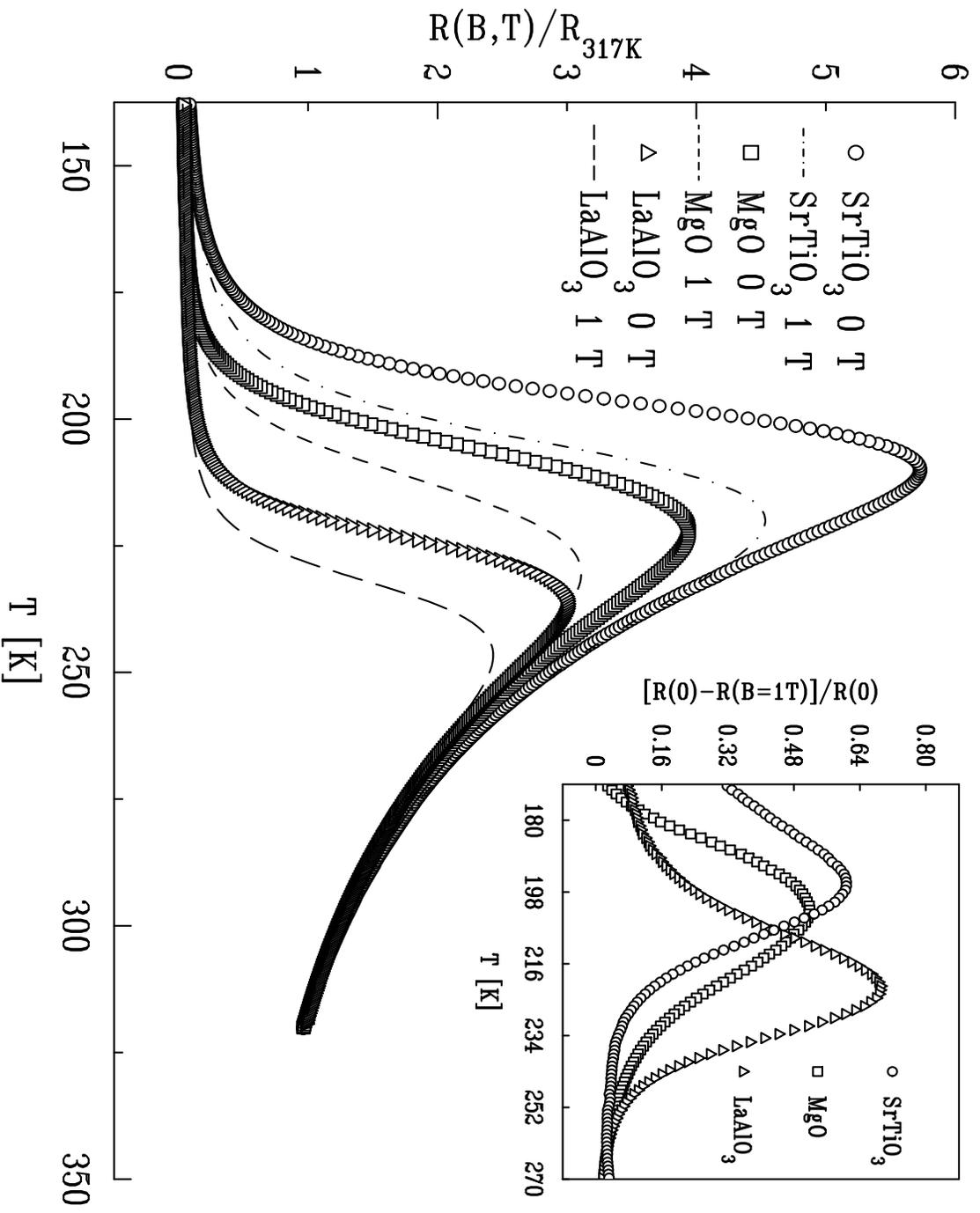
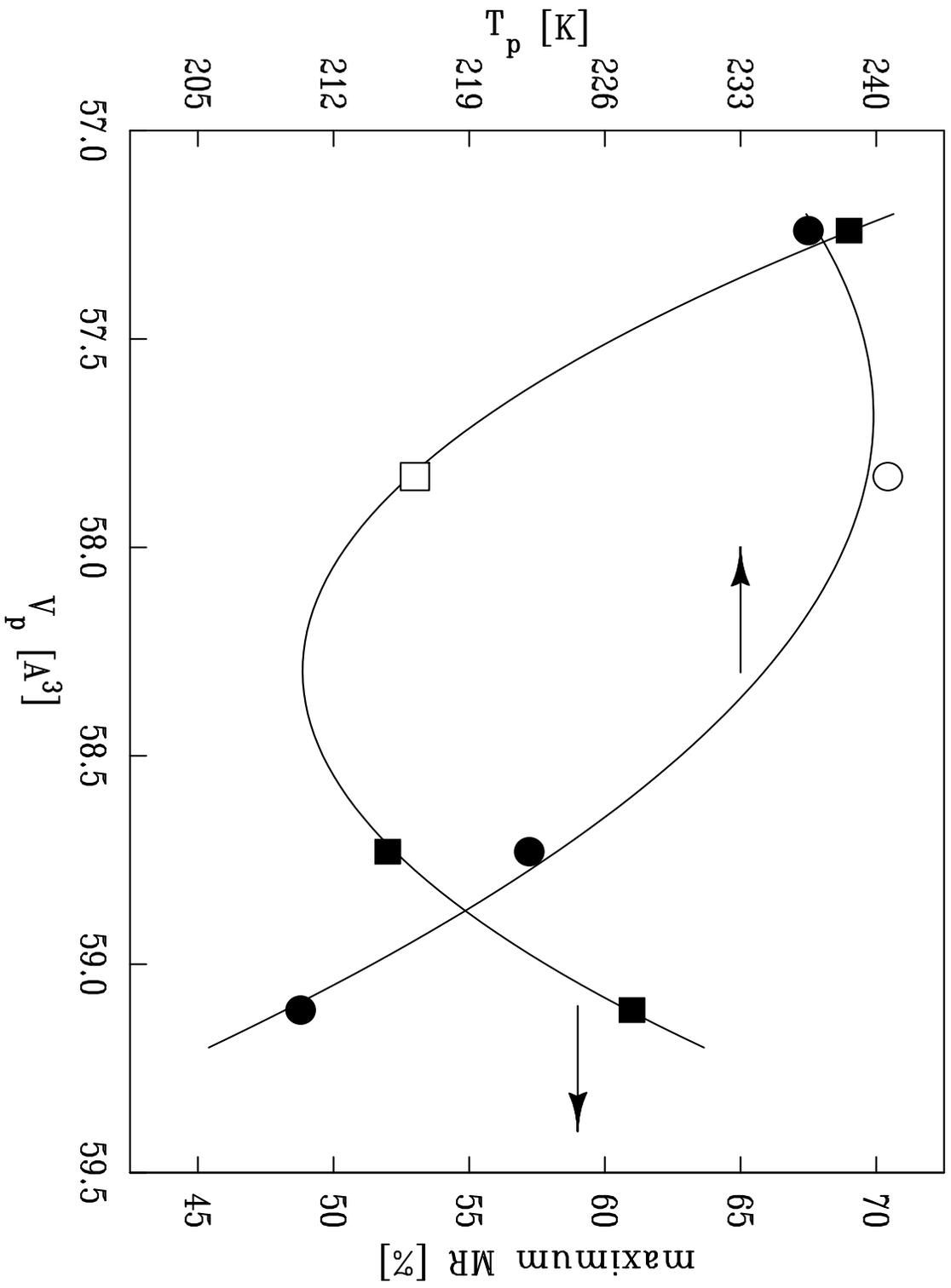
\end{document}